\documentclass[twocolumn]{aastex631}
\usepackage{showyourwork}
\usepackage{amsmath}

\newcommand{\gradrad}{\ensuremath{\nabla_{\rm{rad}}}}
\newcommand{\gradad}{\ensuremath{\nabla_{\rm{ad}}}}

\newcommand{\grad}{\ensuremath{\nabla}}

\begin{document}

\title{Modelling time-dependent convective penetration in 1D stellar evolution}

\author[0000-0002-3054-4135]{Cole Johnston}
\altaffiliation{These authors contributed equally to this work}
\affiliation{Department of Astrophysics/IMAPP, Radboud University, P.O.Box 9010, 6500 GL Nijmegen, The Netherlands}
\affiliation{Institute of Astronomy, KU Leuven, Celestijnenlaan 200D, 3001 Leuven, Belgium}

\author[0000-0001-9097-3655]{Mathias Michielsen}
\altaffiliation{These authors contributed equally to this work}
\affiliation{Institute of Astronomy, KU Leuven, Celestijnenlaan 200D, 3001 Leuven, Belgium}

\author[0000-0002-3433-4733]{Evan H. Anders}
\altaffiliation{These authors contributed equally to this work}
\affiliation{Center for Interdisciplinary Exploration and Research in Astrophysics (CIERA), Northwestern University, 1800 Sherman Ave, Evanston, 60201, IL, USA}
\affiliation{Kavli Institute for Theoretical Physics, University of California, Santa Barbara, CA 93106, USA}

\author[0000-0002-6718-9472]{Mathieu Renzo}
\affiliation{Center for Computational Astrophysics, Flatiron Institute, New York, NY 10010, USA}

\author[0000-0002-8171-8596]{Matteo Cantiello}
\affiliation{Center for Computational Astrophysics, Flatiron Institute, New York, NY 10010, USA}
\affiliation{Department of Astrophysical Sciences, Princeton University, Princeton, NJ 08544, USA}

\author[0000-0002-0338-8181]{P. Marchant}
\affiliation{Institute of Astronomy, KU Leuven, Celestijnenlaan 200D, 3001 Leuven, Belgium}

\author[0000-0003-1012-3031]{Jared A. Goldberg}
\affiliation{Center for Computational Astrophysics, Flatiron Institute, New York, NY 10010, USA}

\author[0000-0002-2522-8605]{Richard H.~D.~Townsend}
\affiliation{Department of Astronomy, University of Wisconsin-Madison, Madison, WI 53706, USA}

\author[0000-0002-7442-1014]{Gautham Sabhahit}
\affiliation{Armagh Observatory and Planetarium, College Hill, Armagh BT61 9DG, N. Ireland}

\author[0000-0001-5048-9973]{Adam S. Jermyn}
\affiliation{Center for Computational Astrophysics, Flatiron Institute, New York, NY 10010, USA}

\correspondingauthor{Cole Johnston}
\email{colej@mpa-garching.mpg.de}
\correspondingauthor{Mathias Michielsen}
\email{mathias.michielsen@kuleuven.be}
\correspondingauthor{Evan Anders}
\email{evanhanders@ucsb.edu}

\begin{abstract}
1D stellar evolution calculations produce uncertain predictions for quantities like the age, core mass, core compactness, and nucleo-synthetic yields; a key source of uncertainty is the modeling of interfaces between regions that are convectively stable and those that are not. Theoretical and numerical work has demonstrated that there should be numerous processes adjacent to the convective boundary that induce chemical and angular momentum transport, as well as modify the thermal structure of the star. One such process is called convective penetration, wherein vigorous convection extends beyond the nominal convective boundary and alters both the composition and thermal structure. In this work, we incorporate the process of convective penetration in stellar evolution calculations using the stellar evolution software instrument {\sc mesa}. We implement convective penetration according to the description presented by \citet{anders_etal_2022a} to calculate a grid of models from the pre main sequence to He core depletion. The extent of the convective penetration zone is self-consistently calculated at each time step without introducing new free parameters. We find both a substantial penetration zone in all models with a convective core and observable differences to global stellar properties such as the luminosity and radius. We preset how the predicted radial extent of the penetration zone scales with the total stellar mass, age and the metallicity of the star. We discuss our results in the context of existing numerical and observational studies.
\end{abstract}

\section{Convective core boundary mixing}
\label{sec:intro}

Stars of all masses and at all stages of evolution rely on convection to transport energy, angular momentum, and chemicals throughout the stellar interior \citep{Lebovitz1965,Aerts2019ARAA,jermyn_etal_2022_atlas}. As such, it is crucial to correctly determine the location where convective transport gives way to radiative transport within the star in order to accurately calculate stellar structure and evolution models \citep{Gabriel2014,Viallet2015}. Convection is an inherently 3D process which typically occurs on very fast time scales compared to most evolutionary time scales \citep[except late burning phases, e.g. silicon burning;][]{Kippenhahn2012}. Therefore, simplified parameterizations of convection are employed in 1D  calculations. The most widely used description of convection in stellar evolution calculations is the Mixing Length Theory (MLT) and its variations and extensions \citep{bohm1958,Spiegel1963,Arnett1969,Canuto1989,Grossman1996,Kupka1996,Gough2001,Salaris2008,Smolec2011,Gough2012,Lesaffre2013,Jermyn2018,Joyce2023}. 

Despite success in broadly reproducing the structure of stars and their evolutionary trajectories, local implementations of convection suffer from an inability to adequately describe the behaviour of convective parcels beyond the convective boundary \citep{Renzini1987,Gabriel2014,Salaris2017,anders_etal_2022b}. Theoretically, there are several phenomena that are expected to occur at and beyond the convective boundary that arise from hydrodynamic properties of convective parcels:
\begin{enumerate}
    \item {\bf Convective penetration} is the extension of vigorous convection beyond the traditional convective boundary, causing the transport and mixing of both chemicals and heat (and therefore entropy) \citep{Shaviv1973,roxburgh_1978,roxburgh_1989,Zahn1991,roxburgh_1992,Viallet2015,Baraffe2023}.
    \item {\bf Convective entrainment} is the process by which convective motions detach from laminar flow and induce turbulent scraping of material from the radiative zone into the convective zone \citep{meakin_arnett_2007,staritsin_2013,scott_etal_2021}.
    \item {\bf Convective overshooting} is the process by which the fluid motions carry a convective cell beyond the convective boundary due to inertia and cause the transport and mixing of chemical elements but do not transport heat across the convective boundary \citep{Saslaw1965,Maeder1975,Stothers1985,Freytag1996,Herwig2000A&A...360..952H}.
    \item {\bf Wave and rotational shear} induced mixing are expected to occur at convective boundaries as well \citep{Mathis2004,Rogers2017,Varghese2023}.
\end{enumerate}   
 Collectively, these mechanisms are referred to as convective boundary mixing (CBM) mechanisms and they are understood to modulate the transport of mass, angular momentum, and chemicals between a convective zone and its adjacent stably stratified radiative zone (see \citet{Anders2023} for a recent review). 

CBM mechanisms have strong implications when a star's nuclear fusing core is convective. In particular, the location of the convective boundary as well as the efficiency of CBM on top of the convective core directly determine the mass of the core and how much nuclear fuel (e.g. Hydrogen, Helium, etc.) is available to undergo fusion \citep{Roxburgh1965,gabriel_etal_2014,Paxton2019,kaiser_etal_2020}. Therefore, CBM directly affects predictions of the masses of stellar cores and luminosities \citep{Kaiser2020,johnston2021,Pedersen2022}, wind mass loss rates \citep{Renzo2017}, stellar ages \citep{Johnston2019b}, the production rates and compactness of stellar end products like neutron stars and black holes \citep{Heger2000,Farmer2016,Davis2019,Farmer2019,Tanikawa2021,Vink2021,Andrassy2022,Mehta2022,Temaj2023}, the strengths of nucleo-synthetic yields \citep{Kaiser2020}, and the properties of stellar populations \citep{Maeder1981,Mermilliod1986,Yang2017,Johnston2019c}. 

Deriving the efficiencies of CBM processes from first principles alone is challenging \citep{Roxburgh1965,Salaris2017}. Instead, when included in 1-D stellar evolutionary calculations, these processes are typically implemented as ad hoc parameterized profiles with a scaled efficiency and a scaled distance over which they are active. Therefore, CBM is treated as a free parameter in evolutionary calculations which is adjusted to match model predictions with observables. 

CBM alters the chemical and thermal gradients near the convective core, therefore modifying the mass of the stellar core as well as observable properties such as the radius and luminosity. In populations of stars, this manifests as an apparent change in the location of the terminal-age main sequence. Thus, both populations of massive O- and B-type field stars and clusters with extended main sequence turn offs have been used to calibrated the relative contributions of rotational mixing and CBM \citep{Mermilliod1986,Brott2011a,Yang2017,Johnston2019c}. 

Since CBM modifies the chemical gradient and core mass of the star, both pressure (p) and gravity (g) mode pulsations are sensitive to the impact of CBM. Numerous studies of pulsating stars have leveraged this to investigate the strength and form of different CBM mechanisms by varying the amount of CBM present in an evolutionary model to match theoretically predicted pulsation frequencies to observed ones in stars \citep{SilvaAguirre2011,Moravveji2015,Buysschaert2018,Szewczuk2018,Johnston2019a,Angelou2020,Deheuvels2020,Viani2020,Wu2020,Noll2021, michielsen_etal_2021,Pedersen2021,Pedersen2022,Michielsen2023}. 

Eclipsing binaries enable the highly precise ($\sim 1\%$) determination of the masses, radii, luminosities, and ages of their components \citep{Torres2010,Maxted2020}. As such, they serve as high fidelity calibrators of CBM processes \citep{Claret2007,Johnston2019b,Serenelli2021}. Crucially, several of these studies have reported that the amount of CBM required to reproduce observations generally scales with the total mass of the star, particularly at lower masses between 1.2-5 M$_{\odot}$ \footnote{It should be noted, however, that there is debate as to the statistical significance of such trends given the precision of modern observations and the degeneracy of free parameters in stellar evolution calculations \citep{Pols1997,Stancliffe2015,Constantino2018,Valle2018,Valle2023}.}  \citep{Andersen1990,Schroder1997,Gimenez1999,Ribas2000,Claret2016,Schmid2016,Valle2018,Claret2019,Angelou2020,Tkachenko2020,Sekaran2021}. An investigation by \citet{Costa2019} attempted to disentangle the contribution of CBM and rotation, arriving at the conclusion that overshooting values can be statistically constrained with modern observations in a limited mass range between 1.5 and 5 M$_{\odot}$. Furthermore, they determine that rotational mixing mechanisms better explain the diversity in chemical mixing values required to match models to observations. \citet{Costa2019} further suggest that using a single mechanism to constrain CBM can often lead to a biased result.

For higher mass stars, a similar trend has been observed for the increase in core mass, which serves as a proxy for the total amount of CBM present in a model \citep{Tkachenko2020}. However, other studies have stressed that varying other parameters, such as the metallicity of the systems or the mixing length value, can account for the apparent trend of CBM with stellar mass \citep{Iwamoto1999,Stancliffe2015,Constantino2018}. When collated, these results reveal that CBM is required for models to match observations and that there is no single value of commonly used CBM free parameters that can reproduce all observations \citep{johnston2021, Anders2023}. 

Recent numerical work has found convective penetration naturally arising in hydrodynamical simulations of convection in stellar interiors \citep{anders_etal_2022a,Baraffe2023,Mao2023,Andrassy2023}. In this work, we study how the structure and evolution of 1D stellar models of intermediate- and high-mass stars is modified by the inclusion of CBM by convective penetration. We implement a convective penetration algorithm which is calibrated from 3D hydrodynamical simulations that calculates the extent of the penetration zone at each time step in the evolution using the properties of the star, i.e. in a self-consistent manner. This implementation differs from previous work by removing free parameters and determining the extent of the penetration zone according to the properties of the model. The result of penetrative convection is a substantial adiabatic region, wherein efficient convective mixing occurs, attached to the core convection zone. Our main result is that we find that the extent of the convective penetration scales strongly with the total mass of the star, in agreement with several notable studies which have observed such a mass dependence.

We find that convective penetration extends the width of the main sequence on the Hertzsprung-Russell diagram (HRD) and substantially alters the properties of stars at the terminal-age main sequence. In Section~\ref{sec:theory} we discuss the supporting theory and implementation within a 1D stellar evolution code (e.g., {\sc mesa}) before presenting the impact on the properties of stars and their evolution in Section~\ref{sec:results}. We discuss the results in the context of the existing literature in Section~\ref{sec:discussion} and conclude in Section~\ref{sec:conclusions}. An implementation of convective penetration routine in {\sc mesa} is publicly available on github\footnote{https://github.com/mathren/dbconvpen}.

\section{Theory and implementation}
\label{sec:theory}

In this work, we implement the dissipation balanced convective penetration algorithm developed by \citet{anders_etal_2022a}. In this section, we briefly describe the theory behind convective penetration and discuss its implementation into a 1D stellar evolution code. Following this 1D implementation, the extent of the penetration zone at any given stage of evolution is determined uniquely by the properties of the star. When this is applied to nuclear burning convective core, this naturally gives rise to a fully deterministic prescription for convective penetration which is stellar mass, age, and metallicity dependent. Below, we briefly describe the theoretical considerations behind our implementation of convective penetration. We note that this implementation can, in practice, be applied to the upper and lower boundaries of both burning and non-burning convective shells as well, however, shells and envelopes are often stratified and therefore the underlying assumptions behind the 3D simulations by \citet{anders_etal_2022a} no longer apply. Furthermore, the simulations did not consider either the partition of buoyant work above and below a convective shell or the entropy gradient at the base of a convective envelope or shell. Therefore, we caution against using this prescription for such cases and we do not consider such application in this work.

\subsection{Theoretical convective penetration}

The convective boundary is defined as the location within the star where the buoyant acceleration vanishes. In the case of a chemically homogeneous medium, this is defined by the Schwarzschild criterion, where the adiabatic and radiative temperature gradients are equal, i.e. $\nabla_{\rm rad}=\nabla_{\rm ad}$. If a chemical gradient is present, however, we adopt the Ledoux criterion such that the boundary occurs where: $\nabla_{\rm rad} = \nabla_{\rm ad} + \nabla_{\mu}$, with the chemical gradient $\nabla_{\mu}= d\ln \mu / d\ln P$, where $\mu$ is the mean molecular weight and $P$ is the pressure. We note that the Ledoux criterion must be applied on the convective side of the boundary since all three gradients are discontinuous at the boundary.

Despite feeling no buoyant acceleration at the convective boundary, convective parcels will still have a velocity and inertia, causing them to travel beyond the boundary. This gives way to the processes listed previously. In this work, we are concerned with convective penetration. As the convective parcels travel beyond the convective boundary, they transport heat as well, effectively mixing entropy and modifying the local temperature gradient such that it approaches $\nabla_{\rm ad}$. This creates an extended convective penetration region above the traditional Schwarzschild (or Ledoux) boundary where $\nabla \sim \nabla_{\rm ad}$. Convective overshooting is still expected to occur above this region (see extended convective penetration described by \citet{Michielsen2019A&A...628A..76M}), but we do not treat this here.

Under the assumption that convection is statistically stationary in time, the buoyant energy generated by convection follows a power integral that can be expressed as:

\begin{equation}
     \int_V F_{\rm conv} \frac{1}{T_0^2}\bigg|\frac{d T_0}{dr}\bigg| dV 
    = \int_V \frac{\Phi}{T_0} dV,
    \label{eqn:roxburgh_2}
\end{equation}
where $F_{\rm conv}$ is the convective flux, $T_0$ is the local temperature in the star, $\Phi$ is the total dissipation (by various mechanisms, e.g. waves), $r$ is the radial coordinate, and $V$ is volume \citep{roxburgh_1989,anders_etal_2022a} . The constraint implies that the buoyant work produced by convection on the left-hand-side of the equation must be balanced out by dissipative forces on the right-hand-side. Following \citet{anders_etal_2022a}, we can express the buoyant engine terms as $\mathcal{B}=F_{\rm conv} \frac{1}{T_0^2}| \frac{dT_0}{dr}|$ and the dissipative terms $\mathcal{D}=\frac{\phi}{T_0}$, and we can break up our integral to consider the contributions to the convective zone (CZ) and penetrative zone (PZ) separately, such that:
\begin{equation}
    \int_{\rm CZ} \, \mathcal{B} dV + \int_{\rm PZ} \mathcal{B} \, dV = \int_{\rm CZ} \, \mathcal{D} dV + \int_{\rm PZ} \mathcal{D} \, dV.
    \label{eq:b_d_equal}
\end{equation}
The PZ is assumed to be fully chemically and thermally mixed such that $\nabla\approx\nabla_{\rm ad}$ despite $\nabla_{\rm rad} < \nabla_{\rm ad}$, therefore, $F_{\rm conv} < 0$ in the PZ. Therefore, we can re-write Eq.~\ref{eq:b_d_equal} such that the positive Buoyant work done by the core is balanced by the positive dissipative and buoyant sinks:
\begin{equation}
        \int_{\rm CZ} \, \mathcal{B} dV  = \int_{\rm CZ} \, \mathcal{D} dV + \int_{\rm PZ} \mathcal{D} \, dV + \int_{\rm PZ} -\mathcal{B} \, dV.
\label{eq:b_d_balanced}      
\end{equation}
In this form, we see that the amount of negative Buoyant work done by the PZ in combination with the dissipative forces in both the PZ and CZ determine the size of the penetration zone required to balance the work produced by the convective zone. Using 3D hydrodynamical simulations, \citet{anders_etal_2022a} simulated convective penetration in the Boussinesq approximation and found that convective penetration is well described by Eq.~\ref{eq:b_d_balanced} re-written as:
\begin{equation}
    -\frac{\int_{\rm PZ} L_{\rm conv} dr}{\int_{\rm CZ} L_{\rm conv} dr} + f\xi \frac{\overline{D}_{\rm PZ}}{\overline{D}_{\rm PZ}} = \left( 1 - f \right),
    \label{eqn:3d_eqn}
\end{equation}
where $L_{\rm conv}$ is the convective luminosity, $\overline{D}$ is the  volume averaged dissipation (in the PZ or CZ), $\xi$ describes the shape of the dissipation profile in the PZ, and f is the ratio of dissipation to buoyant work in the CZ. \citet{anders_etal_2022a} found that their 3D results are well described using $\xi=0.6$ and $f=0.86$.

\subsection{Implementation in 1D evolution code}
In this subsection, we discuss how we port the results from these 3D hydro simulations into a 1D stellar evolution code. This work expands upon previous work by \citet{jermyn_etal_2022} who used a similar method to calculate the extent of the PZ as a post-processing step on the main sequence assuming a chemically homogeneous star. In their work, \citet{jermyn_etal_2022} did not include the effects of the PZ in the evolution of the star. For our implementation, we need a way of estimating the convective luminosity and viscous dissipation rates in the CZ and PZ from the 1D model. Functionally, we balance the following equation (which follows from Eq~\ref{eqn:3d_eqn}):
\begin{equation}
\begin{split}
\int_{\rm PZ} \left[ 4\pi\xi f r^2 F_{\rm avg} + L\left( \frac{\gradad}{\gradrad} - 1 \right) \right] \, dr = \\
(1 - f)\int_{\rm CZ} L_{\rm conv}\, dr,
\end{split}
\label{eqn:functional_eqn}
\end{equation}
where $F_{\rm avg}$ is the volume-averaged luminosity of the convective core, given by:
\begin{equation}
     F_{\rm avg} \equiv \frac{1}{V_{\rm CZ}} \int_{\rm CZ} L_{\rm conv}\, dr.
\label{eqn:F_avg}
\end{equation}
Again, $f$ and $\xi$ are taken from \citet{anders_etal_2022a}. $L$ and $L_{\rm conv}$ are the total luminosity and the luminosity carried by convection, respectively. We estimate the convective luminosity in the penetration zone as: 
\begin{equation}
     -L_{\rm conv} \approx L \left( \frac{\gradad}{\gradrad} - 1 \right).
\label{eqn:L_pen}
\end{equation}
Note that $L_{\rm conv} <0$ in the ${\rm PZ}$ because $\nabla_{\rm rad} < \nabla_{\rm ad}$, and so radiation is carrying the stellar luminosity. We refer the reader to \citet{anders_etal_2022a} for a more detailed description of the simulations and theory supporting this implementation. 

We calculate Eq.~\ref{eqn:functional_eqn} at every time step, and append the PZ on top of the CZ. During a time step, we determine the location of the Schwarzschild boundary of the convective core (i.e., $\nabla_{\rm ad}=\nabla_{\rm rad}$) and store its radial coordinate ($r_{\rm CZ}$) as well as the cell where the boundary occurs. We then integrate the convective luminosity from the center outwards until the boundary cell, accounting for the cases where the boundary occurs within a cell to obtain the right-hand side of the equation. We subsequently integrate outwards from the convective boundary until Eq.~\ref{eqn:functional_eqn} is balanced to determine the boundary of the penetration zone $r_{\rm PZ}$.  For convenience and ease of comparison with typical implementations of CBM mechanisms, we express the extent of the penetration zone as a fraction of the local pressure scale height $h$ evaluated at the Schwarzschild boundary:
\begin{equation}
    \alpha_{\rm pen}=(r_{\rm PZ}-r_{\rm CZ})/h . 
\label{eq:alpha_pen}
\end{equation}
This value $\alpha_{\rm pen}$ multiplied by the local pressure scale height is applied as the extent of the penetration zone above the convective core. We adopt a constant value $D_{\rm pen}$ (${\rm cm\,s^{-2}}$) as the diffusive mixing coefficient across this zone. The exact value of $D_{\rm pen}$ is taken as the diffusive mixing coefficient $f_0$=0.005 pressure scale heights into the convective core. Therefore, $D_{\rm pen}$ is typically of order $D_{\rm conv}$. We further discuss the particulars of an implementation in the 1D code {\sc mesa} in Appendix~\ref{app:mesa}.

\subsubsection{The thermal structure}
\label{sec:thermal_structure}
In a 1D stellar model, we typically enforce that the actual temperature gradient within a star follows the lowest temperature gradient possible. Thus, we expect that beyond the Schwarzschild boundary, where $\nabla_{\rm rad} < \nabla_{\rm ad}$, the temperature gradient will follow the radiative gradient, i.e. $\nabla=\nabla_{\rm rad}$. However, convective penetration is expected to alter the thermodynamic quantities (i.e. the local thermal gradient and entropy) in the PZ \citep{anders_etal_2022a,jermyn_etal_2022_atlas}. Since this behaviour is not naturally captured by MLT, we have to directly modify the temperature gradient such that $\nabla\approx\nabla_{\rm ad}$ within the PZ and then smoothly adjusts to $\nabla_{\rm rad}$. To achieve this, we  follow the implementation of \citet{michielsen_etal_2021} based on the local Pecl{\' e}t number. The Pecl{\' e}t number is a dimensionless quantity that describes the relative importance of convective and radiative heat transport as the ratio of the thermal diffusion time scale to the convective turnover time \citep{jermyn_etal_2022_atlas}. We calculate a local Pecl{\' e}t number, ${\rm Pe}$, based on the local thermal diffusivity, the local mixing length, and the assumed convective velocity that produces the turbulent mixing in the penetration zone; see \citet{michielsen_etal_2021} for more details. We calculate the local thermal diffusivity as:
\begin{equation}
    \chi = \frac{16\sigma T^3}{3\kappa\rho^2 c_p},
\end{equation}
where $\sigma$ is the Boltzmann constant, T is the temperature, $\rho$ is the density, $c_p$ is the specific heat at constant pressure, and $\kappa$ is the Rosseland mean opacity. We then adjust the local temperature gradient as:

\begin{align*}
\label{eq:gradT}
    {\rm Pe>100} \rightarrow \grad&=\gradad,\\
    {\rm 0.01 < Pe < 100} \rightarrow \grad&=k\gradad+(1-k)\gradrad,\\
    {\rm Pe<0.01} \rightarrow \grad&=\gradrad,
\end{align*}
where $k=0.25\left( \log\left({\rm Pe} \right) +2 \right)$.
The selection of 100 and 0.01 as critical numbers is motivated by the behaviour of chemical transport defined by the Pecl{\' e}t number. Values ${\rm Pe}>>1$ are considered adiabatic, whereas values ${\rm Pe}<<1$ are considered diffusive. Therefore, we begin transitioning between adiabatic and diffusive (radiative) transport for values of ${\rm Pe}$ two orders of magnitude larger and smaller than 1. 

For the remainder of this article, we will refer to the extent of the penetration zone as $\alpha_{\rm pen}$, whereas we will refer to the extent of a diffusive step overshooting zone as $\alpha_{\rm ov}$. We make this differentiation to highlight that a penetration zone modifies the temperature gradient, while a traditional step overshooting zone adopts the radiative temperature gradient. 

\section{Convective penetration in 1D evolution models}
\label{sec:results}

\begin{figure*}
    \centering
    \includegraphics[width=0.96\textwidth]{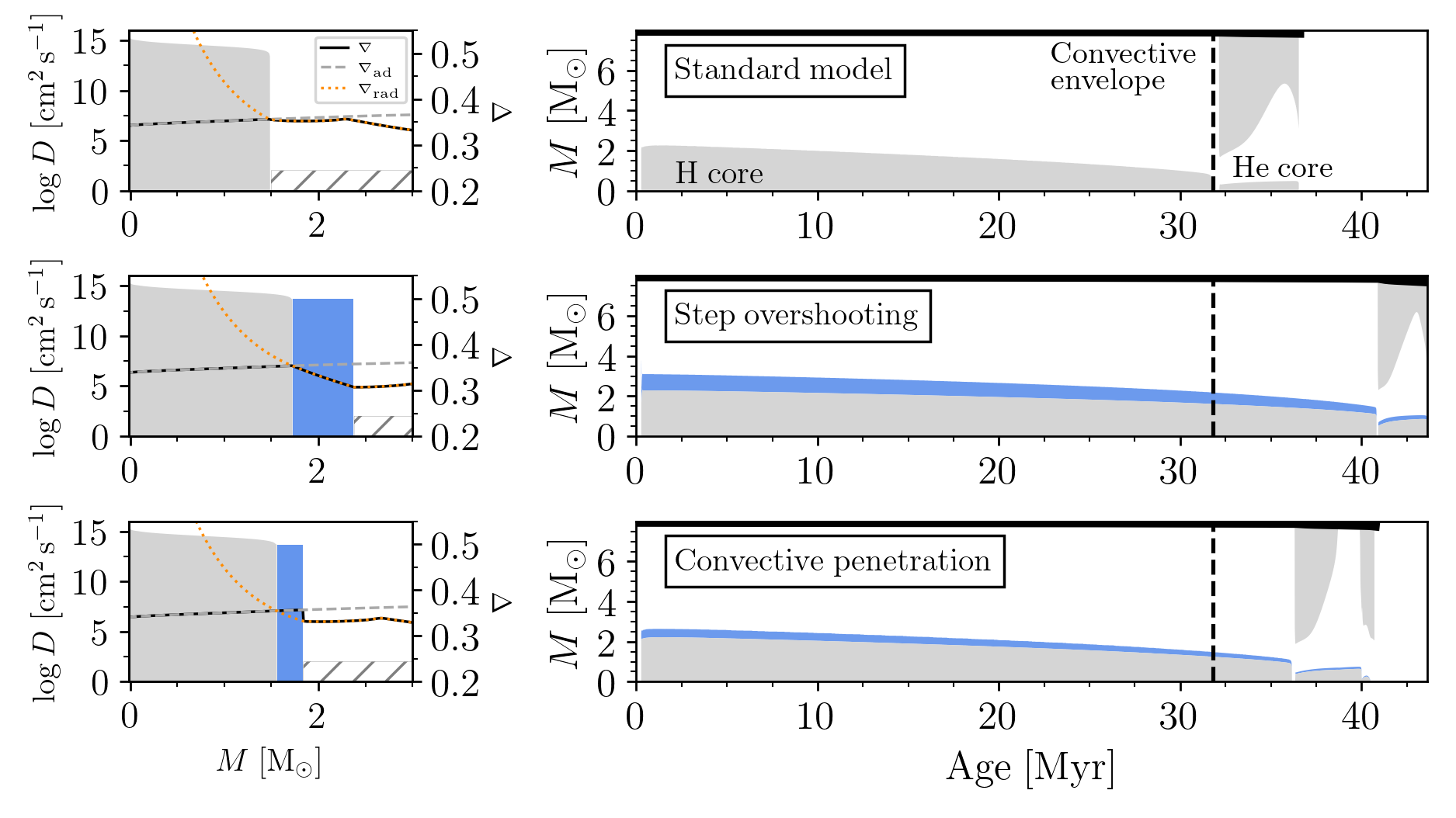}
    \caption{Left column displays the radial mixing profile of the stellar model at $X_c=0.35$ out to 50\% of the stellar mass. Grey regions denote convection, hatched regions denote minimum mixing, and blue regions denote core boundary mixing. The adiabatic, radiative, and actual temperature gradients are shown in black dashed, orange dotted, and black solid lines, respectively. The right column displays Kippenhahn diagrams, with the point of core Hydrogen exhaustion for the standard model with no extra boundary mixing (31.8~Myr) denoted by the vertical dashed black line. The Kippenhahn diagrams use the same color coding as the radial mixing profiles in the left column. The top row is for the model with no convective penetration, and the bottom row is for the model with CBM by convective penetration. For simplicity, the minimum mixing is not shown in the Kippenhahn diagram.}
    \label{fig:kipp}
\end{figure*}

CBM by convective penetration modifies the star in a way that accumulates over the evolutionary history relative to a standard model. In this section, we will examine the results of our study where we implement the above convective penetration algorithm in {\sc mesa} r23.05.1 \citep{Paxton2011,Paxton2013,Paxton2015,Paxton2018,Paxton2019,Jermyn2023}. In this work, we investigate convective penetration on top of convective cores and do not consider convective penetration above or below burning or non-burning convective shells. The details of our {\sc mesa} implementation, including assumptions for various microphysics, are discussed in Appendix~\ref{app:mesa}.

We examine our results from five perspectives. First, we briefly examine how the inclusion of convective penetration alters the structure of the star for an 8 M$_{\odot}$ model. Second, we explore how the extent of the penetration zone depends on stellar mass. Third, we examine how the penetration zone modifies the main sequence evolution of stars of various ages and metallicities. Fourth, we analyze the consequences of including penetrative zones on observables, particularly at the terminal-age main sequence (TAMS). Finally, we analyze the behavior of our implementation for core Helium burning models. For this work, we use a grid of models calculated from $1.2-40$ M$_{\odot}$. Our grid has a variable spacing in mass, such that we more densely sample at lower masses with steps of 0.1 M$_{\odot}$ between 1.2 and 5 M$_{\odot}$, 0.25 M$_{\odot}$ between 5 and 10 M$_{\odot}$, 1 M$_{\odot}$ between 10 and 30 M$_{\odot}$. We do not consider models above this mass as we expect the assumption that the core has a weak density stratification \citep{anders_etal_2022a} to no longer hold above $\sim 50$ M$_{\odot}$ \citep[see Fig.~67 of ][]{jermyn_etal_2022_atlas}. 

\subsection{Structural impact of convective penetration}

As discussed in Section~\ref{sec:theory}, convective penetration effectively extends the region of the star that is adiabatically mixed beyond the traditional Schwarzschild/Ledoux boundary. This results in enhanced chemical transport as well as a modified thermal structure beyond the Schwarzschild/Ledoux boundary. 
Figure~\ref{fig:kipp} displays the structure and evolution of an 8 M$_{\odot}$, Y=0.284, and Z=0.014 model for a model with no CBM (top row), CBM by diffusive step overshooting (middle row) for comparison, and for convective penetration (bottom row). We adopt $\alpha_{\rm ov}=0.33$ for the model with fixed step overshooting to follow the commonly adopted \citet{Brott2011a} value for B-type stars. All models in this example adopt the Ledoux criterion for convective stability and include a minimum amount of diffusive mixing of $D_{\rm min}=100$ cm$^2\,$s$^{-1}$ in the radiative zones. The models displayed in the left column demonstrates the thermal and mixing structure of the models at $X_c=0.35$. The convective zone is denoted by grey and the extent of the CBM zone is shown in blue, while the thermal structure of the star is shown by the orange and black lines. In the top (standard) and middle (step overshooting) models, the thermal structure behaves as expected in 1D models with the convective boundary occurring where $\nabla_{\rm ad}=\nabla_{\rm rad}$. In the bottom (convective penetration) model, however, the actual temperature gradient (solid black line) follows $\nabla_{\rm ad}$ throughout the penetration zone (blue) despite $\nabla_{\rm ad}>\nabla_{\rm rad}$. Finally, we notice that the grey shaded convective core extends to higher masses for the models with step overshooting and convective penetration.

The right panels contain Kippenhahn diagrams showing the evolution of the mixing processes in the stellar interior as the star evolves. The standard 8~M$_{\odot}$ model reaches core Hydrogen exhaustion ($X_c$=$10^{-9}$) at 36 Myr, denoted by the vertical dashed black line in all panels. Comparing these panels, we see that the models including CBM by step overshooting and by convective penetration have more massive cores at all points in time and have longer main-sequence life times. For this 8 M$_{\odot}$ model, adopting the \citet{Brott2011a} step overshooting value of $\alpha_{\rm ov}=0.33$ results in a main sequence that is 28\% longer than the standard model, while the model with convective penetration has a main-sequence lifetime that is 14\% longer than the standard model. Furthermore, using Brott overshooting results in a main-sequence lifetime that is 13\% longer than for the model using our implementation of convective penetration, which has $\alpha_{\rm pen}\approx0.15$.

Beyond the main sequence, the model with convective penetration has a Helium burning phase that is 5\% shorter in total duration than the model with no convective penetration. The decreased duration of the Helium burning phase is caused by a higher central temperature and thus higher energy generation rate that accompanies a more massive Helium core. Additionally, the different fraction of mass held in the stellar core and higher luminosity of the model with convective penetration results in a phase of core-Helium burning where the convective envelope disappears, producing a blue loop (see Section~\ref{sec:he_burn} below). We note that the model with fixed Brott overshooting undergoes a smaller blueward excursion than the model with convective penetration. Finally, examining the Kippenhahn diagram of the model with convective penetration demonstrates that our implementation works robustly for convective cores that are both receding (e.g., on the main sequence) and cores that are  growing (e.g. at the start of core helium burning). 

\subsection{Dependence on stellar mass}
\label{subsec:results_mass_dependence}

\begin{figure*}
    \centering
    \includegraphics[width=0.98\textwidth]{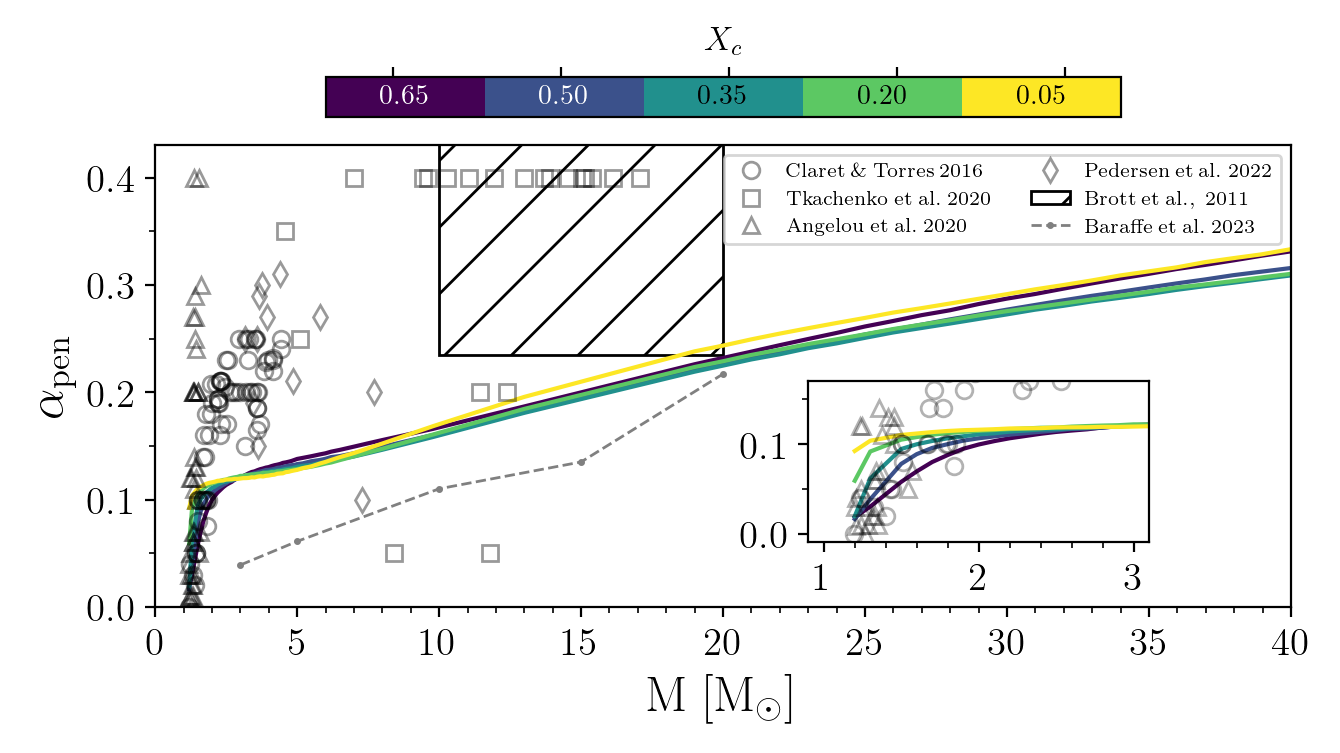}
    \caption{Predicted $\alpha_{\rm pen}$ values as a function of stellar mass and fractional core Hydrogen content ($X_c$). Observationally determined $\alpha_{\rm ov}$ values from various samples in the literature are plotted as well. }
    \label{fig:aov_vs_mass}
\end{figure*}

Following Eq.~\ref{eqn:functional_eqn}, the extent of the penetration zone is largely determined by the amount of luminosity carried by convection, the volume of the convective core, as well as the shape and rate of dissipation beyond the Schwarzschild boundary. The shape and rate of dissipation is described by the terms $\xi$ and $f$, which were calibrated by \citet{anders_etal_2022a}. As both the total luminosity and volume of the convective core increase with stellar mass, we expect the extent of the penetration zone to increase. This is indeed what we find in Figure~\ref{fig:aov_vs_mass} which illustrates mass dependence of the extent of the convective penetration zone $\alpha_{\rm pen}$ on the main sequence. The different colored lines represent different fractional core Hydrogen content values ($X_c$), which functions as a uniform proxy for stellar age during the main sequence. We find a clear monotonic trend with mass. Notably, there is a sharp rise in predicted $\alpha_{\rm pen}$ from 1.2-2 M$_{\odot}$ for models at the ZAMS as seen in the inset of Fig~\ref{fig:aov_vs_mass}. We find a more gradual increase with mass from 2-40 M$_{\odot}$. The trend with age is less obvious to intuit from Eq.~\ref{eqn:functional_eqn}, as the increased convective lumonsity increases towards the end of the main sequence, while the volume of the core decreases. The competition between the two effects generally leads to different trends at different masses. Furthermore, between 1.2-2 M$_{\odot}$, the convective core grows during the first half of the main sequence, leading to an observed increase in $\alpha_{\rm pen}$ with decreasing $X_c$ in this mass range. We note that for models with $2-10$ M$_{\odot}$, the extent of penetration decreases with $X_c$, whereas it increases with $X_c$ for models with $>10$ M$_{\odot}$.

We plot the masses and direct or converted ${\rm \alpha_{ov}}$ values taken from published samples using either binary stars or pulsating stars in grey in Fig.~\ref{fig:aov_vs_mass}. We find a general agreement with the trend that the amount of CBM required to match observations increases with mass. However, we find that our predicted $\alpha_{\rm pen}$ values are systematically lower than those reported by various observational studies for masses $>\sim$2-3 M$_{\odot}$. This is expected as our models only account for convective penetration, while observationally calibrated values of $\alpha_{\rm pen}$ often implicitly include the effects of other CBM mechanisms. We also plot recent predictions of ballistic overshooting from 2D simulations by \citet{Baraffe2023}.

\subsection{Dependence on age and metallicity}

\begin{figure}
    \centering
    \includegraphics[width=0.95\columnwidth]{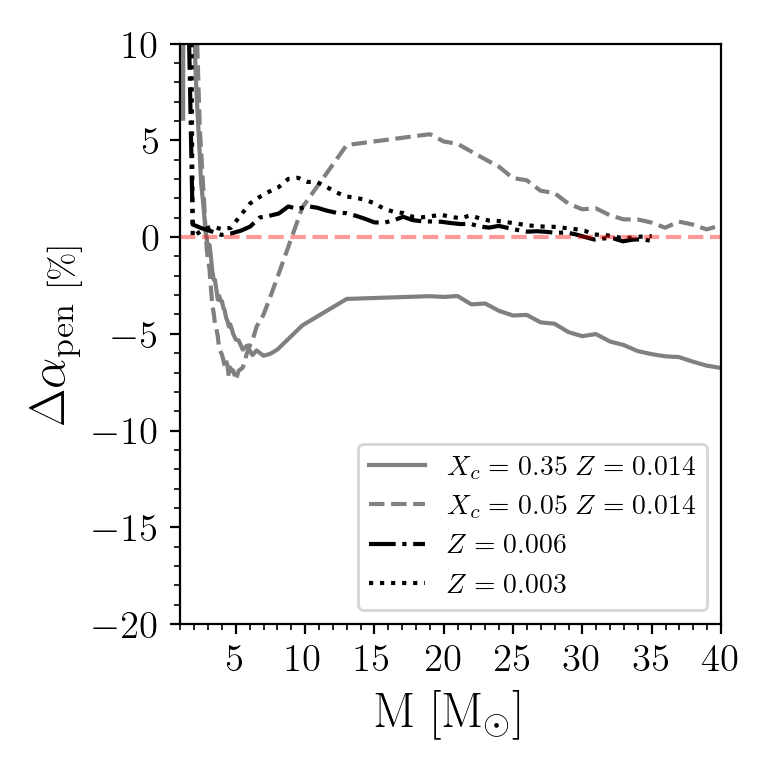}
    \caption{Percent variation in predicted $\alpha_{\rm pen}$ values as a function of mass with respect to $\alpha_{\rm pen}$ for models with $Z=0.014$. The grey line shows the maximum variation in $\alpha_{\rm pen}$ across the main sequence at a given mass with respect to the $\alpha_{\rm pen}$ value at the ZAMS at that mass for models with $Z=0.014$. The difference in predicted $\alpha_{\rm pen}$ at $X_c=0.35$ for models with $Z=0.006$ and $Z=0.003$ with respect to models at $X_c=0.35$ with $Z=0.014$ are shown by the black dash-dotted and black dotted lines, respectively. The horizontal red dashed line denotes zero.}
    \label{fig:aov_variation}
\end{figure}

We find a weak trend with fractional core Hydrogen content, such that older stars have different predicted $\alpha_{\rm pen}$ than their less evolved counterparts. This is seen in the variation in $\alpha_{\rm pen}$ in the different colored lines in Fig.~\ref{fig:aov_vs_mass} and in the solid and dashed grey lines in Fig.~\ref{fig:aov_variation}. We further find a weak dependence of $\alpha_{\rm pen}$ on metallicity at a given mass. These trends are demonstrated in Fig.~\ref{fig:aov_variation}. The solid and dashed grey lines denote the percent difference in $\alpha_{\rm pen}$ at $X_c=0.35$ and $X_c=0.05$ with respect to the $\alpha_{\rm pen}$ at the ZAMS for a fixed mass. Models with masses below 3 M$_{\odot}$ show the largest variation with $\alpha_{\rm pen}$ increasing several tens of percent from ZAMS to TAMS. The maximum decrease in $\alpha_{\rm pen}$ from ZAMS to TAMS is experienced by 5 M$_{\odot}$ stars, whereas models with M $>$ 9 M$_{\odot}$ initially have smaller $\alpha_{\rm pen}$ values towards the middle of the main sequence before producing larger $\alpha_{\rm pen}$ values towards the TAMS. 

The dashed-dotted and dotted black trends depict the percent variation of $\alpha_{\rm pen}$ for models with LMC and SMC metallicities at a fixed evolutionary state ($X_c=0.35$). We see that the dependence with metallicity is strongest at lower masses (i.e. $< 3$~M$_{\odot}$). There is a small variation of $\sim3$\% for models with lower metallicities, however, this variation vanishes at higher masses. The weak dependence on metallicity is likely related difference in the nuclear energy generation rate at lower metallicities in combination with the different total hydrogen content which determines the rate at which the core shrinks on the main sequence as $\kappa\sim(1+X)$ in fully ionized gas \citep{Farrell2022,Xin2022}. Finally, we note that $\alpha_{\rm pen}$ experiences the largest variation in age and metallicity for the lowest mass models which have both a growing and receding convective core on the main sequence, i.e. M$<\sim2.5$ M$_{\odot}$.

\subsection{Observable consequences}
\label{subsec:observations}

\begin{figure}
    \centering
    \includegraphics[width=0.85\columnwidth]{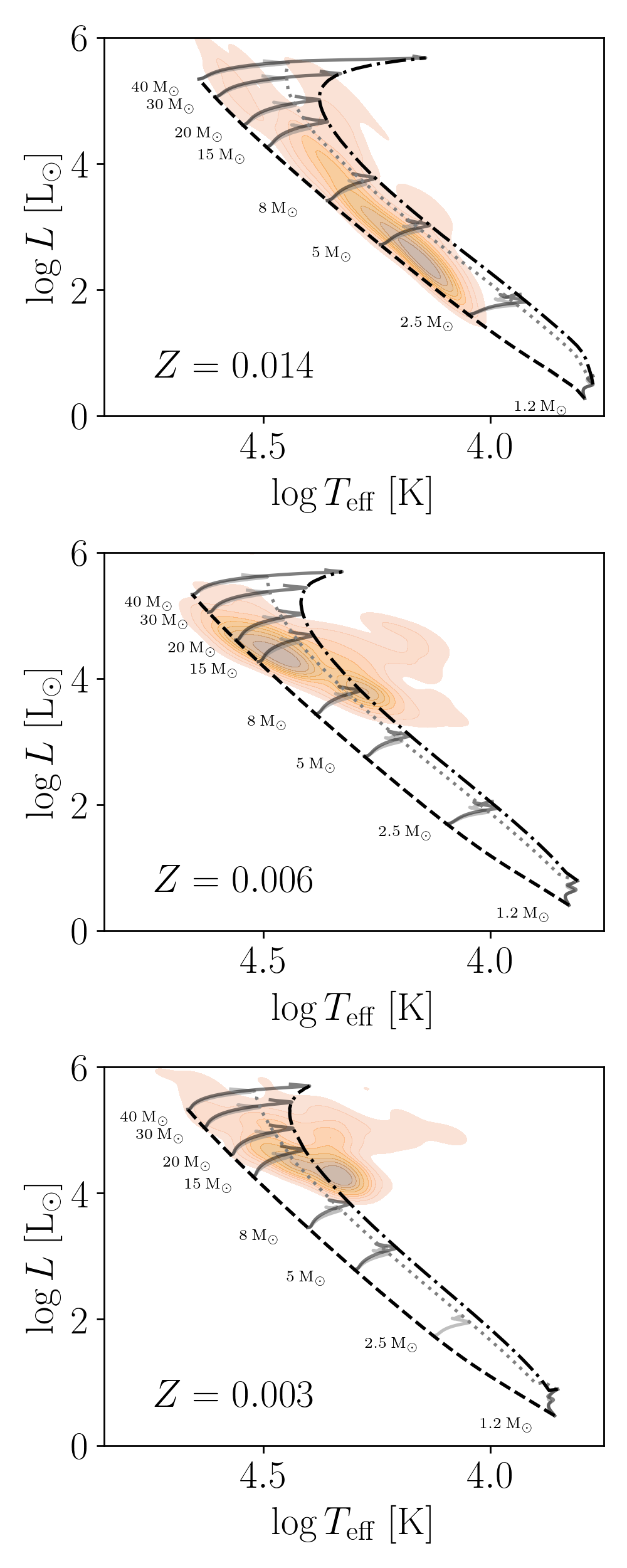}
    \caption{Hertzsprung-Russell diagrams for {\sc mesa} models between 1.2 - 40 M$_{\odot}$ for three metallicity regimes. In all panels, the solid light grey tracks denote models without CBM, while the solid dark grey tracks denote models with CBM by convective penetration. The zero-age main sequence is denoted by a dashed black line, whereas the TAMS is denoted by the dotted light grey line for tracks calculated without CBM, and by the dashed-dotted dark grey line for tracks calculated with CBM by convective penetration. The orange shaded regions are density maps denoting samples of OB type stars in the Galaxy \citep{Dufton2006} and the Magellanic clouds \citep{Hunter2008}.}
    \label{fig:hrd}
\end{figure}

\begin{figure}
    \centering
    \includegraphics[width=0.85\columnwidth]{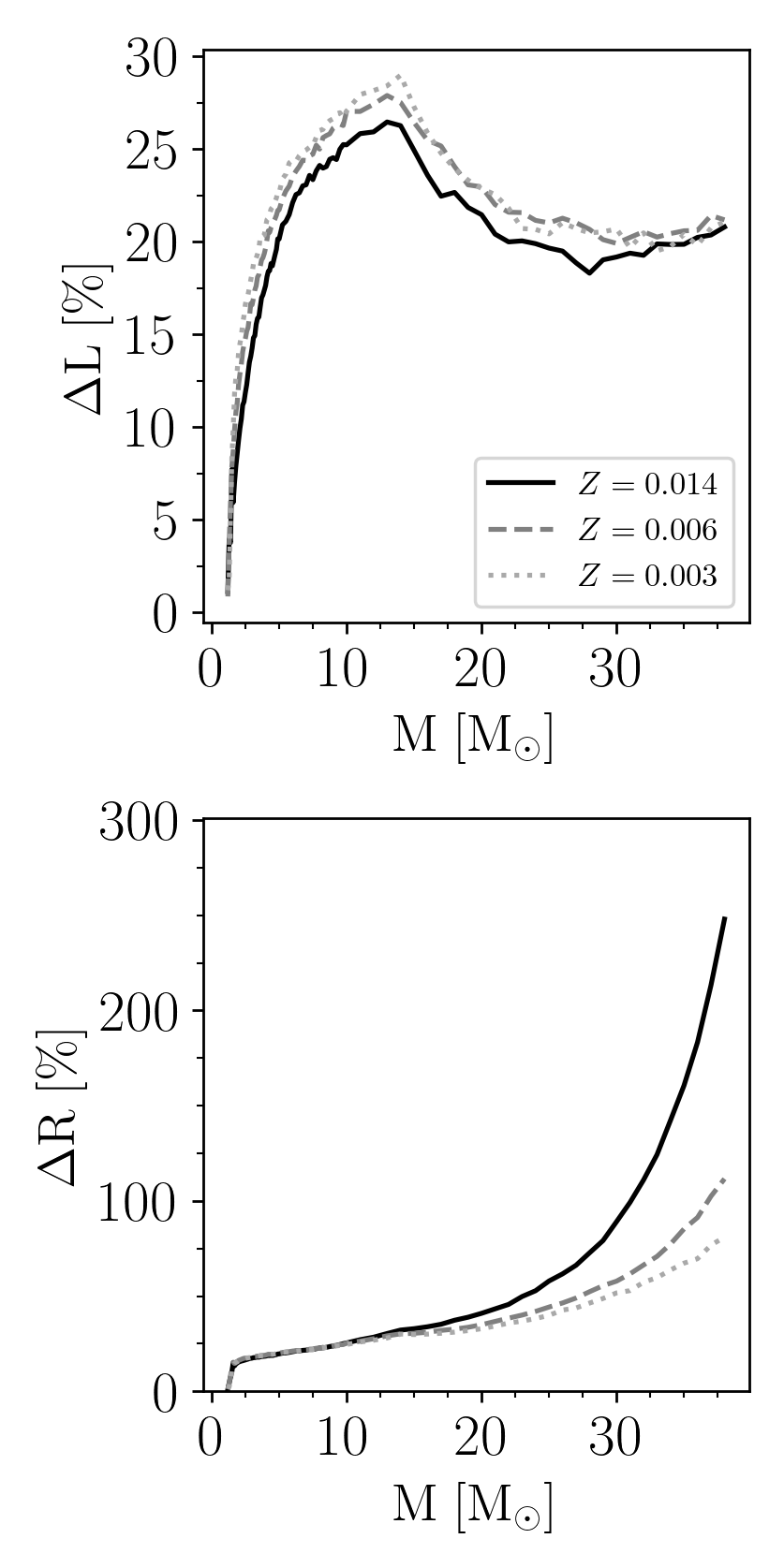}
    \caption{Percent difference in various properties (luminosity, upper left; radius, upper right; age, lower left; remnant Helium core mass, lower right) at the point of core Hydrogen exhaustion between models with and without CBM by convective penetration for three different metallicity regimes.}
    \label{fig:tams_differences_rad_lum}
\end{figure}

During an evolutionary phase with a convective core (e.g., main sequence or He core burning), the convective penetration refuels the burning region and increases the amount of mass held within the convective region, thereby extending the temporal duration of this phase and changing the location of the TAMS in the HRD as the stars evolve to lower surface temperatures before exhausting central hydrogen. This produces a wider main sequence in the HRD. Figure~\ref{fig:hrd} displays the location of the TAMS for models calculated without any CBM (dotted grey lines) and for models with our convective penetration algorithm (dot-dashed black lines) for three different metallicities. The panels of Fig.~\ref{fig:hrd} demonstrate that the inclusion of convective penetration extends the location of the TAMS to cooler temperatures for all metallicities considered. Following the mass dependence of the penetration extent discussed above, we see that the main sequence widens most dramatically for massive stars, following the behaviour of the radii in Fig.~\ref{fig:tams_differences_rad_lum}. The larger displacement of the TAMS for high-mass stars at Z=0.014 is a result of the fact that lower metallicities produce stars will smaller radii, as there is generally less opacity to trap radiation \citep{Kippenhahn2012,Xin2022}. 

The width of the main sequence and location of the TAMS can be observed for populations of stars, and has previously been used to calibrate CBM implementations in stellar models \citep{Maeder1975,Brott2011a,Castro2014,Yang2017,Castro2018,Johnston2019c}. The observed properties of samples of O- and B-type stars in the galaxy, LMC, and SMC are plotted in density maps in Fig.~\ref{fig:hrd}. Based on their inability to match observed samples of Galactic OB stars with models using fixed CBM, \citet{Castro2014} suggested that there was a likely mass-dependence to CBM. Our models with mass-dependent convective penetration manage reproduce the Galactic population of OB stars. However, they fail to match observations of 8-15 M$_{\odot}$ stars in the LMC and SMC samples despite the inclusion of mass-dependent convective penetration. This is not surprising, however, considering our models predict $\alpha_{\rm pen}\approx0.13-0.17$ for these masses, while \citet{Brott2011a} calibrated $\alpha_{\rm ov}=0.335\pm0.1$ for 10-20 M$_{\odot}$ stars in the LMC, indicating that we likely are not accounting for all of the mixing processes in these stars.

The wider main sequence and relocated TAMS is the result of an enhanced core mass and extended core-burning lifetime, which produce changes to the global stellar quantities, such as the overall luminosity and the effective temperature, and thus wind mass loss rates as well. As the impact of convective penetration is cumulative, the changes to stellar properties will be most apparent and most observable at the end of the convective core burning phase considered (e.g., TAMS, end of He core burning). We plot the predicted difference between the luminosity and radius of models calculated with and without convective penetration for three metallicities in the panels of Fig.~\ref{fig:tams_differences_rad_lum}. The differences are calculated as a percentage of the value for the standard model at the TAMS, e.g. $\Delta= 100\times \left(L_{\rm conv.~pen.}-L_{\rm standard}\right)/L_{\rm standard}$. The models with convective penetration reach a maximum difference of being $\sim$25-30\% more luminous than models without convective penetration at $\sim15$ M$_{\odot}$, but have little dependence on the metallicity. In contrast, higher metallicity stars with convective penetration have increasingly larger radii at the TAMS. We note that the more massive models are beginning to experience inflated radii as they approach their Eddington limit \citep{Sanyal2015}. Additionally, these models have started evolving into the Hertzsprung gap, resulting in an extended and no longer easily identifiable TAMS.

\begin{figure}
    \centering
    \includegraphics[width=0.85\columnwidth]{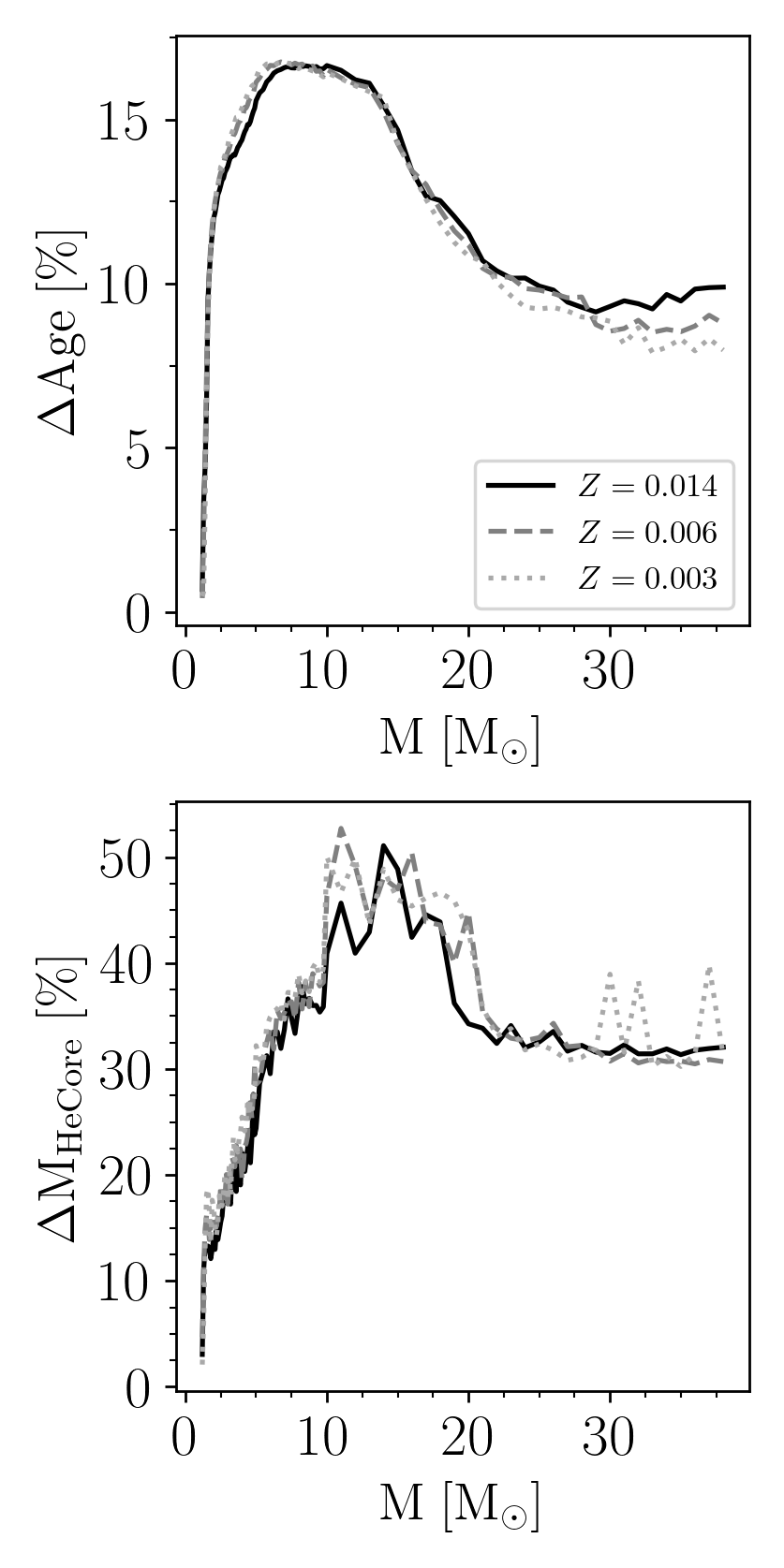}
    \caption{Percent difference in various properties (luminosity, upper left; radius, upper right; age, lower left; remnant Helium core mass, lower right) at the point of core Hydrogen exhaustion between models with and without CBM by convective penetration for three different metallicity regimes.}
    \label{fig:tams_differences_age_cmass}
\end{figure}

The convective penetration zone results in both a longer main-sequence lifetime and a more massive convective core at any stage of evolution. Therefore, models including convective penetration will have a more massive remnant Helium core at Hydrogen exhaustion than models calculated with no CBM. The panels of Fig.~\ref{fig:tams_differences_age_cmass} display the difference in age and remnant Helium core mass for models calculated with and without convective penetration. The inclusion of convective penetration in evolutionary calculations produces a mass dependent difference in the age and remnant Helium core mass difference at the TAMS. Our models predict that convective penetration has a lesser impact on the age and Helium core mass at the TAMS for stars above $\sim$20 M$_{\odot}$ compared to stars at lower masses. All models display an increased helium core mass at the TAMS, with the largest difference of nearly $50$\% occurring for stars with $\sim15$ M$_{\odot}$. This difference in remnant helium core mass will have important consequences for the later evolution of stars, in particular for the duration of later core burning stages, as well as the rates of supernovae, neutron star and black hole production \citep{Fryer2012,Farmer2019,Patton2020,scott_etal_2021,Zapartas2021,Fryer2022,Temaj2023}.

\subsection{Core helium burning models}
\label{sec:he_burn}

\begin{figure}
    \centering
    \includegraphics[width=0.98\columnwidth]{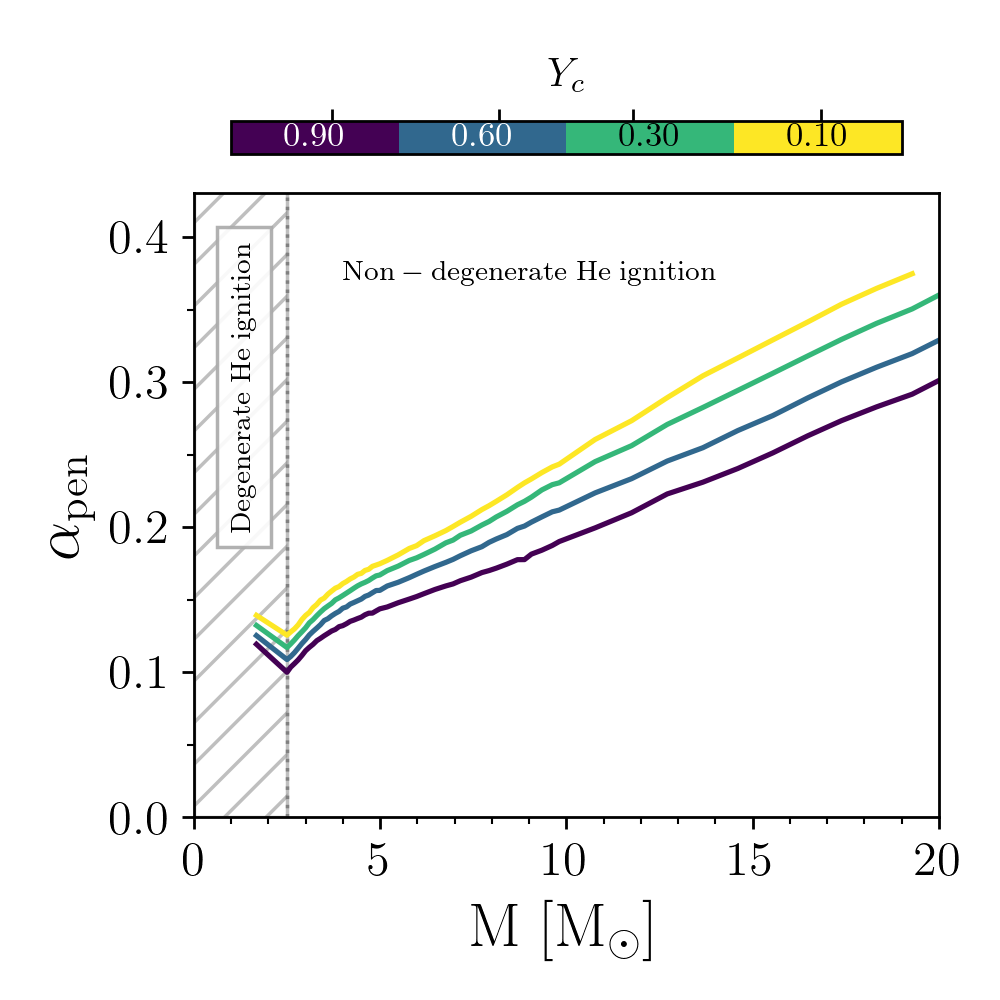}
    \caption{Predicted $\alpha_{\rm pen}$ values as a function of stellar mass and fractional core Helium content ($Y_c$).}
    \label{fig:aov_vs_mass_cHe}
\end{figure}

While only stars with M $>$ 1.2 M$_{\odot}$ have a convective core on the main sequence, all subsequent core burning stages exhibit convective cores. Thus, the inclusion of (physically motivated) CBM in later burning stages is of great importance to stellar evolution calculations, and in particular the final fates of stars. Specifically, there has been observational work estimating the role of CBM in red giant stars and more massive Cepheid variables \citep{Keller2008,Bossini2015}. Thus, we evolved models from 1.5 to 20 M$_{\odot}$ to core Helium exhaustion and plot their predicted $\alpha_{\rm pen}$ values as a function of mass and core Helium fraction in Fig.~\ref{fig:aov_vs_mass_cHe}. Similar to the case of models on the main sequence, we find a strong mass dependence of the predicted $\alpha_{\rm pen}$ with mass. We further find a stronger dependence with fraction core Helium content than models on the main sequence, with models predicting larger $\alpha_{\rm pen}$ values as they approach core Helium exhaustion. We note that the inversion at $2.5$ M$_{\odot}$ is the result of models transitioning from igniting Helium under degenerate conditions to non-degenerate conditions, as annotated in Fig~\ref{fig:aov_vs_mass_cHe}. 

\begin{figure}
    \centering
    \includegraphics[width=0.98\columnwidth]{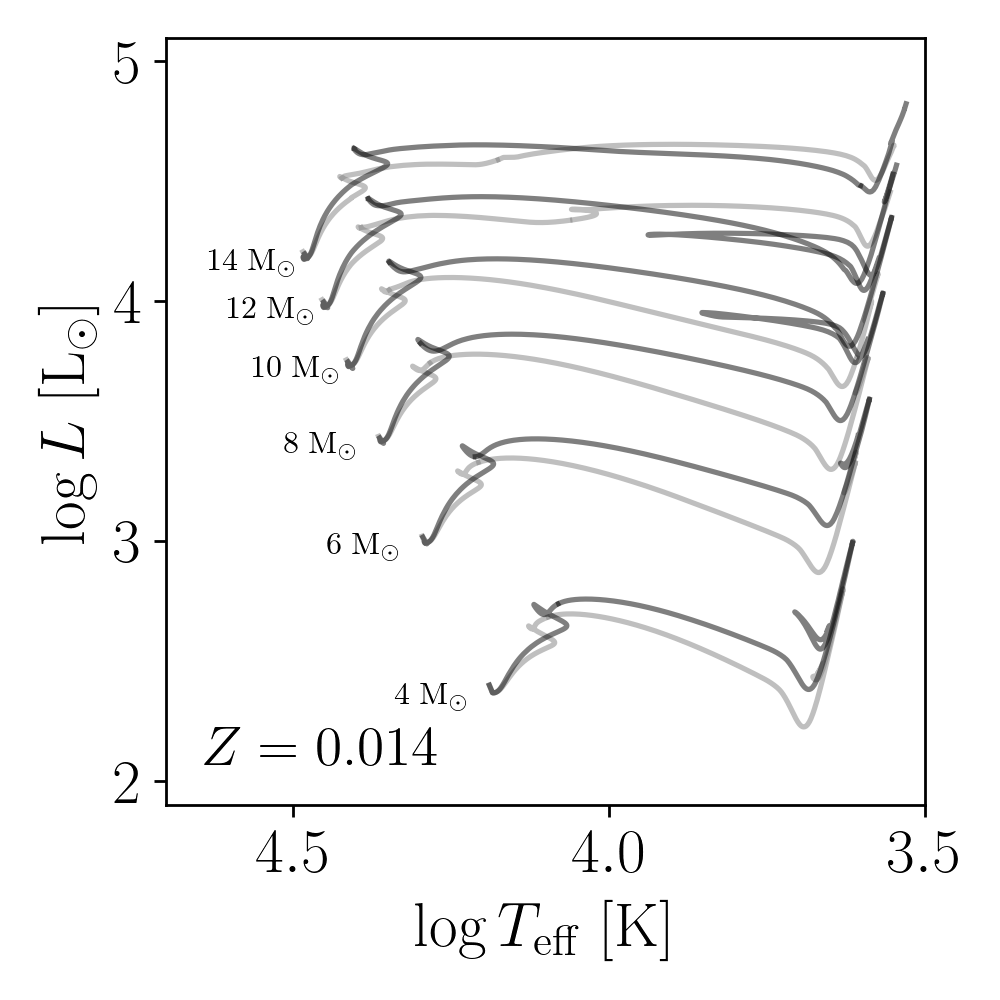}
    \caption{Evolutionary tracks for models with M=4,6,8,10,12,14 M$_{\odot}$ from zero-age main sequence to the end of core Helium burning. Light grey tracks represent models with no CBM, and dark grey tracks represent models with CBM by convective penetration.}
    \label{fig:cepheid_hrd}
\end{figure}

Red giant and supergiant models sometimes perform loop-like blueward excursions during the core He burning phase, termed as "blue loops". The occurrence of and mechanism behind blue loops have been the topic of extensive academic discussion and debate over the last half century \citep{Schlesinger1977,Stothers1979,Renzini1992,Walmswell2015}. While the existence of objects such as pulsating Cepheid variables indicates that blue loops should occur, whether or not a blue loop occurs in 1D models depends on the H and He gradients above the core \citep{Schlesinger1977,Stothers1991,Walmswell2015,Farrell2022}. As the inclusion of convective penetration in our models change these gradients both on the main sequence and beyond, we investigate the behavior of stars in the Cepheid mass range with our models. We plot the full evolutionary tracks of models between 4-14 M$_{\odot}$ in Fig.~\ref{fig:cepheid_hrd} for models without any CBM in light grey and for models with convective penetration in dark grey. The models with $10$ M$_{\odot}$ and $12$ M$_{\odot}$ clearly demonstrate large blue loops, while the 4 M$_{\odot}$ displays a smaller blue loop. Conversely, none of the models without CBM display a blue loop in this mass range. We note that the jagged behavior of the light grey standard 12 M$_{\odot}$ and 14 M$_{\odot}$ models in the Hertzsprung gap is likely caused by the absence of any CBM in our models, which leads to sharper discontinuities in the composition gradients near the convective boundary. This can cause jumps in the convective boundary location as determined by the Ledoux criterion and amplify other (often numerical) uncertainties. Finally, we note that we do not consider penetration above or below convective envelopes or shells, which would also modify the H and He gradients responsible for producing blue loops \citep{Farrell2022}.

\section{Discussion}
\label{sec:discussion}

The main result of this work is the natural emergence of the mass-metallicity-age dependence of $\alpha_{\rm pen}$ when we self-consistently implement convective penetration as calibrated by 3D simulations in a 1D stellar evolution code. This result exists within a broader context of many other numerical and observational studies surrounding the role of CBM in stellar evolution. In this section, we discuss how our predictions for $\alpha_{\rm pen}$ as a function of mass, metallicity, and age compare to the estimates obtained by various numerical and observational studies. We further discuss the implications that our results have for stellar evolution beyond core H exhaustion.

\subsection{Comparison with numerical work}
As seen in Fig.~\ref{fig:aov_vs_mass}, our implementation predicts $\alpha_{\rm pen}=0.05-0.3$ (at Galactic metallicity), depending on the mass of the star. This result is in {\it general} agreement with recent hydrodynamical simulations from various teams who investigate convective penetration at different masses. Specifically, we note that the general mass dependence of $\alpha_{\rm pen}$ agrees with the results of \citet[][2D and in-compressible equations]{Higl2021} and \citet[][2D and fully compressible equations]{Baraffe2023} who have independently demonstrated that the amount of CBM present in hydrodynamical models should scale with total stellar mass. We note that while our predicted $\alpha_{\rm pen}$ values share a similar trend as those reported by \citet{Baraffe2023}, we predict values nearly a factor two larger than found by \citet{Baraffe2023}. This difference is likely caused by the different considerations of what to measure, where \citet{Baraffe2023} consider the maximum distance of a ballistic convective cell while we consider the distance needed to dissipate the buoyant work generated by convective flows above the Schwarzschild boundary. Furthermore, our results agree with the recent work by \citet[][2.5D and 3D with fully compressible equations]{Andrassy2023} who demonstrated that convective penetration naturally arises in hydrodynamical simulations of a 15~M$_{\odot}$ main-sequence star with an extent ranging from $\alpha_{\rm pen}=0.09-0.44$, depending on the approximations used (our models predict $\alpha_{\rm pen}=\sim0.18$ for a 15~M$_{\odot}$ main-sequence star). We comment that our results are derived from simulations that only concern mixing induced by convective penetration under the Boussinesq approximation, whereas the work of other studies use different approximations, geometries, and numerical techniques. 

Recently, work by \citet{kupka_etal_2022} and \citet{Ahlborn2022} have successfully implemented two versions of a model for turbulent convection \citep{Kuhfuss1986} in a 1D stellar evolution code. This model successfully accounts for the non-local behavior of convective parcels, and predicts non-zero convective velocities and adiabatic mixing above the Schwarzschild boundary \citep{Ahlborn2022}. While both their models and ours have adiabatic mixing ($\nabla\sim\nabla_{\rm ad}$) beyond the Schwarzschild boundary, the exact profile of $\nabla$ from our models does not exactly match predictions from either of their models (compare our Fig.~\ref{fig:kipp} with their Fig.~3). Further comparison becomes difficult as the authors do not differentiate between the effects of convective penetration ($\nabla\sim\nabla_{\rm ad}$) and mechanical overshooting ($\nabla\rightarrow\nabla_{\rm rad}$) in their results. Future work comparing the penetration region between these two approaches could help to further understand the behavior of 1D implementations of convection. 

\subsection{Comparison with observations}

In recent decades, there have been numerous studies that have tried to calibrate different implementations of CBM using samples of pulsating or binary stars, or distributions of temperatures and surface gravities in populations. Generally, our models predict lower values of $\alpha_{\rm pen}$ than are estimated in these samples. Even at low masses, we predict a smaller penetration extent than was observationally determined by \citet{Claret2019} who related their results to same integral constraint introduced by \citet{roxburgh_1992}. The systematic difference is due to how CBM values are observationally calibrated. Most studies include a single CBM mechanism in their evolution models, with a few notable exceptions \citep{Brott2011a,Costa2019,Higgins2019}. Thus, when the models are matched to the observations, the reported CBM values act as a proxy for all CBM mechanisms and not a single mechanism. In this way, we expect our predicted $\alpha_{\rm pen}$ values, which only account for mixing by convective penetration, to be systematically lower than observational $\alpha_{\rm ov}$ values. Even though they included other mixing mechanisms, \citet{Costa2019} still report a wide range of $\alpha_{\rm ov}=0.3-0.8$ in stars with 1.5 M$_{\odot}$ $<$ 5 M$_{\odot}$.

The location of the TAMS is often used to calibrate CBM in evolution models \citep{Maeder1981,Brott2011a,Castro2014,Yang2017}. Notably, \citet{Brott2011a} used a sample of OB stars in the Large Magellanic cloud (LMC) to calibrate $\alpha_{\rm ov}=0.33\pm0.1$ for stars between 10-20 M$_{\odot}$, as depicted by the black-hatched region in Fig.~\ref{fig:aov_vs_mass}. Similarly, \citet{Rosenfield2017} and \citet{Yang2017} find that large values of $\alpha_{\rm ov}=0.5$ and $\alpha_{\rm ov}=0.4$ are required for their models to match observations of the width of the extended main-sequence turn off in young massive clusters of stars. Similar to the case of samples of binaries and pulsating stars, our models predict $\alpha_{\rm pen}$ for stars in these mass ranges that are lower by a factor $\sim2$. This again could be due to the fact that our predicted $\alpha_{\rm pen}$ values consider a single CBM mechanism, while observationally calibrated values serve as a proxy for the total chemical mixing history of the star. Furthermore, \citet{Castro2014,Castro2018} use samples of OB stars in the Galaxy and LMC and SMC to suggest a mass-dependent amount of CBM. While they do not suggest an exact amount, our models predict this mass-dependence. 

Evolutionary models often require higher input masses to match the observed effective temperatures, luminosities, and surface gravities than those masses that are observationally determined. This is generally referred to as the `mass discrepancy' \citep{Herrero1992}. Numerous mechanisms have been proposed to reduce the mass discrepancy, such as rapid rotation and gravity darkening, wind-driven mass loss, magnetic fields, opacities, sub-surface convective zones, and improved atmospheric models \citep{Stothers1991,Weidner2010,Brott2011a,Castro2014,Markova2018}. \citet{Higgins2019} discussed the competing roles that rotation, CBM via overshooting, and mass loss play in matching evolutionary models to observations, demonstrating that multiple mechanisms are required to remedy the mass discrepancy. Recently, \citet{Tkachenko2020} demonstrated that the mass discrepancy can be reduced by considering models with a more massive convective core (for a fixed stellar mass). However, many of the models require $\alpha_{\rm ov}$=0.4 to match the observed properties of the stars. 

Generally, there is a wide diversity of reported $\alpha_{\rm ov}$ values required to match models to observations. While the observational literature broadly indicates that there is a trend between increasing stellar mass and CBM levels, there is a large scatter in the reported values \citep{johnston2021}. This is likely caused by varying methodologies and is compounded by degeneracies between stellar evolution model parameters \citep{Aerts2018,Johnston2019b}. There are even numerous studies that report that no extra CBM is required to match models to observations, or that the amount of CBM cannot be constrained by modern observations and techniques, signalling the difficulty in addressing this issue \citep{Pols1997,Stancliffe2015,Constantino2015,Valle2018}. The benefit of our implementation is that it reduces the dimensionality of the modelling procedure by removing the free parameter that controls the extent of the penetration zone. This does not, however, address the full range of mixing processes that are expected at the convective boundary.

\subsection{Later burning stages \& future evolution}
\label{sec:future_evol}

While the majority of the literature surrounding calibrating CBM has focused on the Hydrogen burning main sequence, several notable studies have demonstrated the importance of correctly determining the convective boundary and the role of CBM in core Helium burning stars \citep{Schwarzschild1970,Castellani1971,Ostrowski2017,Constantino2017,Li2021}. In particular, \citet{Constantino2015} and \citet{Bossini2015} demonstrated that CBM can help minimize apparent discrepancy in observed and theoretically calculated period spacing values for mixed modes in red clump stars \citep{Montalban2013}. \citet{Bossini2015} further demonstrate that the asteroseismic signature of mixed modes in core Helium burning stars indicate the need for a CBM region with an adiabatic temperature stratification. Notably, these studies rely on models with fixed amounts of CBM for all masses, with mixed success. Following these results, we note that our prescription for convective penetration with both a mass and evolution dependent $\alpha_{\rm pen}$ value and a modified adiabatic temperature gradient stands as a good candidate to address the systematic offsets between observed and predicted pulsation properties of core Helium burning giants.

At higher masses, whether cool supergiant models stabilize as a red supergiant or experience a blue loop is highly sensitive to the chemical gradient above the Helium core, and any processes that modify it, such as CBM, efficient semi-convection or a deep dredge-up episode \citep{Stothers1968,Langer1985,Farrell2022}. Cepheid variable stars, which traverse the Hertzsprung gap, display a discrepancy between the masses derived from their pulsational properties and the masses required as input into evolutionary codes to reproduce their observed spectroscopic properties \citep{Stobie1969,Keller2008,PradaMoroni2012}. Alongside updated opacities and improved mass loss prescriptions, CBM has been proposed as a solution to mediate the observed discrepancy \citep{Keller2008,Neilson2011}. Studies investigating the role of CBM in reducing the Cepheid mass discrepancy have indicated that large values of step overshooting ($\alpha_{\rm ov}>0.3$) are required to reproduce observations \citep{Neilson2012,Miller2020}. Following Fig.~\ref{fig:aov_vs_mass_cHe}, our models only predict $\alpha_{\rm pen}$ values above 0.3 in models of M$>$14 M$_{\odot}$. We further note that the natural emergence of blue loops (Fig.~\ref{fig:cepheid_hrd}) indicates that convective penetration can help mitigate some of the Cepheid mass discrepancy. By consistently including convective penetration at all stages of evolution, this will further reduce the mass discrepancy while requiring smaller amounts of CBM during the core Helium burning stage alone.

The ratio of red to blue supergiants (R2) in clusters has been used as a means of calibrating CBM. \citet{Constantino2017} demonstrated that CBM by overshooting impacts R2, but find values that are much smaller ($f_{\rm ov}=0.001\rightarrow\alpha_{\rm ov}\approx0.01$) than are predicted by our models. \citet{Constantino2018} implemented a CBM scheme that modulates the rate of ingestion of Helium into the core according to the local luminosity and temperature gradient above the convective core. While direct comparison with this model is challenging, they find high rates of ingestion that suppress breathing pulses are required to match observations of R2 from Galactic globular clusters.

Apart from R2, another important feature concerning the upper HRD is the general absence of observed cool supergiants above a certain luminosity threshold \citep{Humphreys1979}. While mass loss is generally considered as the primary mechanism that sets this limit, multiple studies have also pointed out the importance of internal mixing \citep{Maeder1982,Higgins2020,Gilkis2021,Sabhahit2021}. Following these implications, future work will systematically investigate the influence of penetrative mixing on the post-MS evolution of massive stars, mainly on the blue-to-red supergiant ratio and the luminosity threshold of cool supergiants. Finally, convective penetration will naturally produce more massive remnant Carbon/Oxygen cores at the end of Helium burning and affect their composition \citep{Farmer2019,Laplace2021,Farmer2023}, with important consequences on the `explodability' of the final stellar structure \citep{Ertl2016,Sukhbold2016,Patton2020,Zapartas2021,Patton2022}. A detailed investigation into the impact that this has on the production of neutron stars and black holes is highly valuable, but beyond the scope of this paper.

\section{Conclusions}
\label{sec:conclusions}

In this work, we presented {\sc mesa} stellar evolution models which incorporate a self-consistent implementation of convective penetration which has been calibrated from 3D hydrodynamical simulations by \citet{anders_etal_2022a}. We calculated the extent of the penetration zone at each time step of the evolution, accounting for the structure of the star. To explore the impact that the inclusion of convective penetration has on stellar structure and evolution, we calculated grids of models between 1.2-40 M$_{\odot}$ at several metallicities. 

Using these grids, we found that the predicted $\alpha_{\rm pen}$ values in our models strongly depend on the total mass of the star, and weakly depend on the evolutionary stage and metallicity of the star during the main-sequence evolution, as displayed in Figs.~\ref{fig:aov_vs_mass} and \ref{fig:aov_variation}, respectively. We found that the models with convective penetration produce observable differences in various stellar quantities, such as the luminosity, radius, age, and Helium core mass, as demonstrated in Figs.\ref{fig:tams_differences_rad_lum},~\ref{fig:hrd}, and \ref{fig:tams_differences_age_cmass}. We further found that our models predict that $\alpha_{\rm pen}$ depends strongly on the total stellar mass and fractional core Helium content in core Helium burning stars as seen in Fig.~\ref{fig:aov_vs_mass_cHe}. By considering the impact of convective penetration through the end of core Helium burning, we showed that some models naturally experience blue loops, as seen in Fig.~\ref{fig:cepheid_hrd}. 

Although the predictions of models generally agree with various estimates of the effects of CBM on the main sequence in the literature, we caution against making direct comparisons to results from specific studies. In practice, the task of estimating the efficiency or extent of a given CBM mechanism from observations is extremely difficult and often suffers from implicit degeneracies in stellar model parameters that cannot be fully mitigated given the precision of modern observations and modelling techniques \citep{Aerts2018}. Furthermore, estimates for the extent of one CBM mechanism are often conflated estimates of the total CBM present in a stellar model, leading to difficulties in comparing our results which strictly account for the effects of CBM by convective penetration. However, our results do support the argument that (at least some forms of) CBM display a dependence on stellar properties, most strongly with stellar mass. 

In summary, we have isolated the impact of convective penetration on stellar evolution over a wide mass range and over 99\% of the stellar lifetime ($\sim90$\% core H burning, $\sim10$\% core He burning). This work highlights the importance of including mass-dependent CBM in stellar evolution calculations, and provides a means to do so in a deterministic manner. With this work, we isolated the role of convective penetration as a CBM mechanism. This will enable future studies to isolate the role of other CBM mechanisms such as mechanical overshooting, mixing via internal gravity waves, and rotationally induced shear mixing in stellar evolution. In future work, we plan to address the role of mechanical overshooting, and explore the ability of detailed asteroseismic and binary modelling to scrutinise this implementation of convective penetration. Finally, future work will also seek to improve upon the 3D hydrodynamical simulations upon which this work was based to include the effects of radiation pressure and density stratification, allowing us to apply this prescription to masses above 40~M$_{\odot}$.

\section*{Acknowledgments}
The authors thank the anonymous referee for their comments which improved the quality and clarity of the manuscript. CJ, MM, and EHA are are grateful for the kind hospitality offered by the staff of the Center for Computational Astrophysics at the Flatiron Institute of the Simons Foundation in New York City 
during their work visit in the fall of 2022. The Center for Computational Astrophysics at the Flatiron Institute is 
supported by the Simons Foundation. CJ, MM, and EHA thank Conny Aerts for her discussions about stellar structure and evolution. EHA further thanks Daniel Lecoanet for many discussions about CBM. CJ additionally thanks Isabelle Baraffe for discussions about CBM and multi-dimensional simulations. CJ gratefully acknowledges support from the Netherlands Research School of Astronomy (NOVA) and from the Research Foundation Flanders (FWO) under grant agreement G0A2917N (BlackGEM). The research leading to these results has received funding from the Research Foundation Flanders (FWO) by means of a scholarship to MM under project No. 11F7120N. EHA was supported by a CIERA Postdoctoral Fellowship and by a KITP Postdoctoral Scholar position, so this research was supported in part by the National Science Foundation under Grants No.~NSF PHY-1748958 and PHY-2309135. PM acknowledges support from the FWO senior fellowship number 12ZY523N. RHDT acknowledges support from NASA grant 80NSSC20K0515. The resources and services used in this work were provided by the VSC (Flemish Supercomputer Center), funded by the Research Foundation - Flanders (FWO) and the Flemish Government. Some computations were conducted with support by the NASA High End Computing Pro- gram through the NASA Advanced Supercomputing (NAS) Division at Ames Research Center on Pleiades with allocation GIDs s2276, so this work was also supported in part by NASA HTMS grant 80NSSC20K1280.

\section*{Software} This work has made use of: {\sc mesa} \citep{Paxton2011,Paxton2013,Paxton2015,Paxton2018,Paxton2019,Jermyn2023}, the {\sc mesa}SDK \citep[version 22.6.1; https://doi.org/10.5281/zenodo.7457681, ][]{townsend2022_mesasdk}, as well as the {\sc numpy} \citep{Harris2020}, {\sc 
matplotlib} \citep{Hunter2007}, {\sc scipy} \citep{SciPy2020} and {\sc pandas} \citep{Pandas2010,Pandas2020}.

\appendix

\section{MESA implementation}
\label{app:mesa}

Our 1D stellar evolution calculations were performed using the Modules for Experiments in Stellar Astrophysics software instrument \citep[{\sc mesa} \texttt{r23.05.1}]{Paxton2011, Paxton2013, Paxton2015, Paxton2018, Paxton2019, Jermyn2023}. We have also tested this implementation on {\sc mesa} \texttt{r22.11.1} We implement both the convective penetration algorithm and the accompanying modification to the thermal structure of star using the \texttt{other} hook scheme in {\sc mesa}. We note that we take care to appropriately calculate weighted averages when combining both cell center and cell face quantities in a calculation. 

In order to consistently account for the structure of the star in determining the size of the convective penetration zone, we employ the use of the \texttt{ other\_overshooting\_scheme} hook in {\sc mesa}. During a time step, we determine the location of the Schwarzschild boundary of the convective core (i.e., $\nabla_{\rm ad}=\nabla_{\rm rad}$) and store its radial coordinate ($r_{\rm CZ}$) as well as the cell where the boundary occurs. We then integrate the convective luminosity from the center outwards until the boundary cell, accounting for the cases where the boundary occurs within a cell to obtain the right-hand side of the equation. We subsequently integrate outwards from the convective boundary until Eq.~\ref{eqn:functional_eqn} is balanced, then determine the boundary of the penetration zone $r_{\rm PZ}$. The extent of the penetration zone is calculated as a fraction of the local pressure scale height $h$. This value $\alpha_{\rm pen}$ is then passed to {\sc mesa}, where a diffusive mixing profile with value $D_{\rm pen}$ (${\rm cm\,s^{-2}})$ is applied to the model with a step profile. This profile extends $\alpha_{\rm pen}$ local pressure scale heights above the convective core. The value $D_{\rm pen}$ is taken as the diffusive mixing value at $f_0=0.005$ pressure scale heights into the convective core. Varying $f_0$ would have a small but largely negligible effect on the results. Making $f_0$ larger/smaller would slightly increase/decrease the diffusive mixing coefficient throughout the penetration zone, but would not change the extent of the penetration zone. Finally, to include the impact of the modified thermodynamic quantities on the stellar structure we use {\sc mesa}'s \texttt{other\_adjust\_mlt\_gradT\_fraction} hook, scaling the temperature gradient according to the Pecl{\' e}t number, as described in Section~\ref{sec:thermal_structure}. 

In this work, we calculate the extent of the convective penetration zone $\alpha_{\rm pen}$ according to the radial (Eulerian) coordinates of the convective core and penetration zone. This approach is stable in our calculations, however, in phases of rapid evolution (i.e. core Hydrogen/Helium exhaustion) the radial extent of the convective core and penetration zone can change rapidly compared to their mass coordinate. Furthermore, this implementation was calibrated using a Lagrangian coordinate system in the 3D hydrodynamical simulations by \citet{anders_etal_2022a}. To be consistent, we developed a separate implementation of the convective penetration algorithm that uses the mass coordinate to determine the extent of the convective penetration zone. Following our tests, we find the same results from both implementations.

\subsection{Input physics}
\label{subsec:mesa_stock}

For our models, we adopt several fixed parameters. We adopt the Cox MLT description and set $\alpha_{\rm MLT}=1.8$. We set a minimum diffusive mixing coefficient of $D_{\rm min}=100$ ${\rm cm\,s^{-1}}$ in all models. We use non-rotating models and use the `T\_tau' atmosphere with the Krishna-Swamy relation. We adopt the `Dutch' wind scheme \citep{Glebbeek2009} with a scaling factor of 0.8. All of our models use mesh\_delta\_coeff=0.4 and time\_delta\_coeff=0.5, however we ran convergence tests to demonstrate that this does not modify our results. 

The {\sc mesa} EOS is a blend of the OPAL \citep{Rogers2002}, SCVH \citep{Saumon1995}, FreeEOS \citep{Irwin2004}, HELM \citep{Timmes2000}, PC \citep{Potekhin2010}, and Skye \citep{Jermyn2021} EOSes.

Radiative opacities are primarily from OPAL \citep{Iglesias1993,Iglesias1996}, with low-temperature data from \citet{Ferguson2005} and the high-temperature, Compton-scattering dominated regime by \citet{Poutanen2017}. 
Electron conduction opacities are from \citet{Cassisi2007} and \citet{Blouin2020}.

Nuclear reaction rates are from JINA REACLIB \citep{Cyburt2010}, NACRE \citep{Angulo1999} and additional tabulated weak reaction rates \citet{Fuller1985, Oda1994, Langanke2000}. 
Screening is included via the prescription of \citet{Chugunov2007}.
Thermal neutrino loss rates are from \citet{Itoh1996}.

For models at the Galactic metallicity, we adopt $Z=0.014$ and $Y=0.284$ in accordance with the Galactic B star standard \citep{Nieva2012}. When calculating models at different metallicities, we determined the Helium abundance using the typical enrichment law $Y=Y_p+Z(\Delta Y/\Delta Z)$, where $Y_p=0.2451$ and $(\Delta Y/\Delta Z)=2$ \citep{Pols1998,Valerdi2019}.

\bibliography{johnston_accepted}

\end{document}